# A Concept For Elimination Of Small Orbital Debris

By Gurudas GANGULI[1], Christopher CRABTREE[1], Leonid RUDAKOV[2], and Scott CHAPPIE[3]

[1] Code 6756, Plasma Physics Division, Naval Research Laboratory, Washington DC 20375-5346
[2] Icarus Research Inc., P.O. Box 30780, Bethesda, MD 20824-0780
[3] Code 8223, Naval Center For Space Technology Naval Research Laboratory, Washington DC 20375-5346

A concept for forced reentry of small orbital debris with characteristic dimension ~ 10 cm or less from the highly populated sun synchronous orbit by injecting micron scale dust grains to artificially enhance drag is discussed. The drag enhancement is most effective when dust grains counter rotate with respect to the debris resulting in hypervelocity dust/debris impacts. While the natural drag on small debris with ballistic coefficient ~ 5 kg/m$^2$ in orbits with perigee above 900 km is negligible, it is sufficient to decay the orbit of the injected dust grains at a significant rate. This offers a unique opportunity to synchronize the rates of descent of the dust and debris to create a sweeping (snow-plow-like) effect on the debris by a descending narrow dust layer. The dust density necessary to de-orbit small debris is sufficiently low such that the orbits of active satellites which have larger ballistic coefficients are minimally affected. If deemed necessary, contact of the injected dust with active satellites may be avoided by maneuvering the satellite around the narrow dust layer.

## 1. Introduction

Space debris can be broadly classified into two categories: (i) large debris with dimension larger than 10 cm and (ii) small debris with dimension smaller than 10 cm. The smaller debris are more numerous and are difficult to detect and impossible to individually track. This makes them more dangerous than the fewer larger debris which can be tracked and hence avoided. In addition, there are solutions for larger debris, for example, NRL's FREND device that can remove large objects from useful orbits and place them in graveyard orbits[1]. To the best of our knowledge there are no credible solutions for the small debris. Damage from even millimeter size debris can be dangerous. Fig. 1 shows examples of damage by small debris collision. The source of small debris is thought to be collision between large objects[2], such as spent satellites, which can lead to a collisional cascade[3]. Perhaps a more ominous source of smaller debris is collision between large and small objects as we describe in the following. Since such collisions will be more frequent our focus is to develop a concept for eliminating the small orbital debris which can not be individually tracked to evade collision.

## 2. Small Debris Population

The LEO debris population is primarily localized within a 50 degree inclination angle and mostly in the sun synchronous nearly circular orbits[4]. The distribution of larger trackable debris peaks around 800 km altitude. The smaller debris, although impossible to track individually, can be characterized statistically[5] and the resulting distribution is roughly similar to the tracked debris but peaks at higher (~ 1000 km) altitude. The lifetimes of debris increase with their ballistic coefficient, $B$, defined as the ratio of mass to area[6]. Debris with $B \sim 3-5$ kg/m$^2$ peak around 1000 km and their lifetime becomes 25 years or less below 900 km. Above 900 km the lifetimes can be centuries. Therefore, the task of small debris removal is essentially to reduce the debris orbit height from around 1100 km to below 900 km and then let nature take its course. Today there are about 900 active satellites and about 19,000 Earth-orbiting cataloged objects larger than 10 cm. However, there are countless smaller objects that can not be tracked individually. Unintentional (collision or explosion) or intentional (ASAT event) fragmentation of satellites increases the debris population significantly. For example, the 2007 Chinese ASAT test generated 2400 pieces of large debris and countless smaller ones in the popular sun synchronous orbit at 900 km altitude[7]. A similar increase of the debris population also resulted from the 2009 collision of the Iridium 33 satellite with a spent Russian satellite Kosmos-2251. These collisions are examples of high energy fragmentation where the energy dissipated is several hundreds if not thousands of MJ and the average velocity spread of the fragments $\Delta V$ could be several hundred m/s. Since the population of smaller debris ~ 10 cm size is at least an order of magnitude higher, their collision frequency with larger objects would correspondingly be an order of magnitude higher. However the energy in such collisions is typically less than 10 MJ.

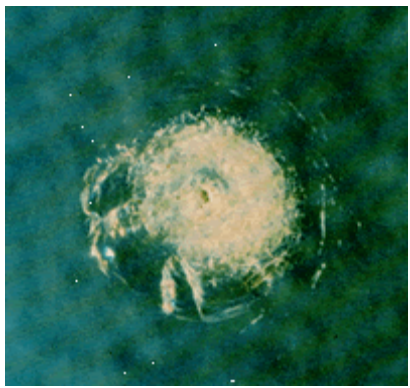

Fig. 1a. A 4 mm diameter crater on the windshield of the space shuttle created by impact of 0.2 mm paint chip.

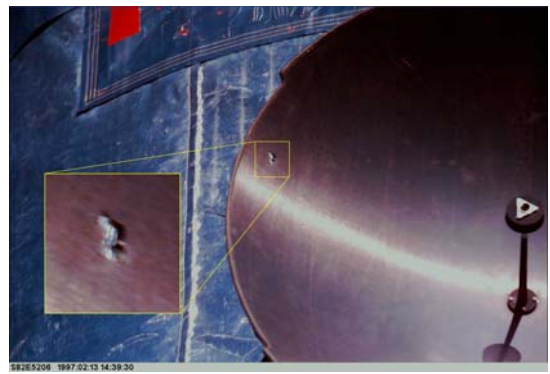

Fig. 1b. Impact of less than 1 cm debris created a hole on an antenna of the Hubble space telescope[13].

### 3. Low Energy Fragmentation

Collision of small debris with large objects could also create secondary small debris. To understand fragmentation in a low energy collision we make simple scaling arguments to the NASA high energy fragmentation model[8]. Consider a debris fragment, such as a piece of satellite structure (Al, size ~10 cm, 30-50 g), which collides with a satellite weighing 500 kg and about a meter in characteristic size. Considering a relative velocity of 15 km/s the debris kinetic energy is about 3-5 MJ, which is equivalent to the explosive power of about 1 kg TNT. The collision is likely to puncture a hole in the satellite external structure (as in Fig. 1b), break apart the internal structures of the satellite into smaller pieces, and increase the pressure inside the satellite. Since 3 MJ spread over $1m^3$ is equivalent to 30 atmospheres, the satellite structure would be subjected to 10-30 atmospheres from inside. Under such a jump of pressure the satellite will break up and small fragments generated by the impact inside the satellite would be expelled out as secondary small debris. Such break up of satellites may not be as catastrophic as the high energy fragmentation and hence $\Delta V$ of fragments is expected to be much smaller than that which would result from a high energy fragmentation. We can estimate $\Delta V$ of the expelled fragments by scaling to the well studied case of the Chinese ASAT test. According to Johnson et al. (2008) the Fengyun-1C was destroyed by a ballistic kinetic kill vehicle (KKV) which collided with the satellite with a relative velocity of approximately 9 km/s. Assuming the mass of the KKV to be between 50 - 80 kg the kinetic energy is about 2000 MJ which is about 600 times larger than the kinetic energy delivered by a typical small debris. The $\Delta V$ of the fragments from Fengyun-1C can be estimated to be around 300 m/s[8,9]. Since the kinetic energy is proportional to $V^2$ we expect the debris fragments from a collision with a typical small debris to have a velocity that is $\sqrt{600} \approx 25$ times less; i.e., for a typical low energy fragmentation $\Delta V \sim 10\,m/s$. Clearly, the characteristic of the low energy fragmentation is quite different from the high energy fragmentation[8] but it can generate secondary small debris. Therefore, removal of small orbital debris is just as, if not more, important than the removal of larger objects because they are also a source for secondary small debris and due to larger population their collision frequency is much higher.

### 4. Physics Basis for Small Debris Elimination

In this article we only consider the case where debris has uniformly spread around the earth, i.e., fragmentation has occurred long before the remediating dust is applied. The altitude of small debris can be reduced to below 900 km by artificially increasing the drag on the debris. Higher drag can be achieved by injecting dust grains in a similar but oppositely directed orbit with respect to the targeted debris. Due to orbit perturbations caused by the Earth's irregular gravitational field, the orbital debris and dust orbits will precess. However, injection in nearly polar orbits will minimize dust precession. Dust would spread over latitude due to spread in the injection velocity and finally form a narrow shell slowly spreading in azimuth with dust meridianal velocity in both directions. As the debris population enters this shell during the course of their precession they experience enhanced drag. At any point in time half of the debris population entering the dust shell will be counter rotating with respect to the dust.

The drag on the debris orbit is determined by the momentum balance equation,

$$\underbrace{\left(\frac{M}{A}\right)}_{Ballistic\ Coefficient} \frac{dV}{dt} = -\underbrace{n_d m_d}_{Dust} \times \underbrace{(V - v_d)^2}_{Relative\ Velocity} \times \kappa - \frac{C_D}{2} \underbrace{n_o m_o}_{Atmosphere} V^2, \quad (1)$$

where M, A, V, are debris mass, area (exposed to drag), and velocity, while $m_d$, $v_d$, and $n_d$, are the mass, velocity, and density of dust. $C_D$ is the coefficient for atmospheric drag and the factor $\kappa = (1 + \sqrt{1+f})$ in Eq. (1) accounts for the type of dust/debris collision. If $f = -1$ then it implies the dust is stuck on the debris on impact i.e., inelastic collision, $f = 0$ implies elastic collision, and $f > 0$ implies loss of debris mass as ejecta due to melting or evaporation resulting from hypervelocity impacts.

Maximum drag is achieved when the relative velocity between dust and the debris, $V_{rel} = V - v_d = 2V$ where $V \approx 7.5$ km/s is the orbital speed. This implies hypervelocity dust-debris collision at about 15 km/s which will result in debris evaporation and melting and effectively increase the drag force by a factor $\kappa$, where $fm_d$ is the debris mass that melts at hypervelocity impact with dust grain. For a specific example assume the debris to be aluminum with specific heat of melting of 0.35 KJ/g. The specific kinetic energy of tungsten dust grains at 15 km/s is 110 kJ/g. Due to the shock generated at hypervelocity impact most of the dust kinetic energy will be used to melt and vaporize debris[10] and debris mass of about $300m_d$ will melt per impact. This corresponds to $f = 300$ and hence $\kappa \sim 18$. This value of $\kappa$ is conservative because additional ejecta mass due to fragmentation in the micro-crater generated by the hypervelocity impact is not considered. This will result in larger $\kappa$ which implies lower dust mass as indicated in the estimate of dust mass given in Eq. (2) below.

In our baseline estimate we use $\kappa = 18$ based on the assumption that ejecta (i.e., thrust) is formed due only to melting of the debris material. Accurate value of $\kappa$ is not known but is necessary for reliable estimate of dust mass necessary for the dust based Active Debris Removal (ADR) system that we propose.

## 5. Dust Mass Estimate

The natural atmospheric drag is included in the second term in the right hand side (RHS) of Eq. (1) through atmospheric density $n_0$ and mass $m_0$. At higher altitudes of interest, e.g., 900 – 1100 km where small debris population is high, the atmospheric drag on the debris is negligible and hence their orbital lifetime is very long. Their lifetime can be shortened to the extent desired by artificially enhancing the drag through the injection of dust, represented in the first term in RHS of Eq. (1). However the atmospheric drag on 30 – 70 $\mu$m diameter dust grains is not negligible and hence the dust orbit will naturally decay. The dust orbit decay rate is dependent on the dust grain size and mass density and hence, to a certain extent, can be controlled. We can exploit this by injecting a narrow dust layer of width $\Delta R$ which is much smaller than the altitude interval $\delta R$ to be cleared (see Fig. 2) and synchronizing the rate of descent of the debris and the dust. As the dust descends in altitude due to the natural atmospheric drag, it 'snow plows' the small debris until a low enough altitude is reached below which the natural drag is strong enough to force reentry of the debris. Since $\Delta R << \delta R$ the volume of dust is much less than the volume of the interval to be cleared. Hence, the dust mass to be transported to orbit can be kept at a minimum. In addition, small $\Delta R$ (30 – 50 km) allows for the option to maneuver active satellites to avoid contact with the injected dust.

Consider the case in which the debris orbit altitude is to be lowered by $\delta R$ below which the natural drag is sufficient to reduce the lifetime of the debris to a desired interval. Neglecting the second term it can be shown from Eq. (1) that the total dust mass $M_d$ necessary for this is,

$$M_d \sim \frac{\delta R \Delta R}{4\kappa NC} B, \qquad (2)$$

where N is the number of debris revolutions in the dust, which is a measure of dust/debris interaction time. In LEO the period of the debris revolution is about 90 min which implies that there are about 5200 revolutions a year. $C \sim (0.5 - 1)$ is a correction factor due to the orbital geometry and assumed to be 0.5 in the following. In deriving (2) we have used $\Delta v / v = \delta R / 2R$. From Eq. (2) $M_d$ necessary to lower the orbit heights of all $B \leq 5$ debris from 1100 km to below 900 km in 10 years by releasing 60 $\mu$m diameter tungsten dust in a layer of width $\Delta R \sim 30$ km at 1100 km is estimated to be 20 tons, which is about 1 m$^3$ of dust. This corresponds to $10^{13}$ grains of dust in a volume $\sim 10^{25}$ cc resulting in dust number density $\sim 10^{-12}$/cc. The dust may be injected in one or several installments over a period of several years. For this estimate we assumed $\kappa = 18$. This value is likely to be larger which implies that the estimate of the dust mass is likely to be lower. Further research is necessary to determine a more accurate value of $\kappa$.

The length of time required to de-orbit the small debris is influenced by how long we can maintain the dust in orbit. In the near-earth plasma environment the dust grains acquire charges and respond to the electromagnetic forces in addition to gravity, drag, and radiation pressure depending on its size and composition. Orbit calculations using silicon and tungsten dust of a variety of sizes from 1-100 $\mu$m indicate that the orbital lifetime of dust depends on its size and density. The on/off radiation pressure due to dust orbit in sunlight and in earth shadow introduces a spatial spread to its Keplerian orbit. These calculations suggest that 30 - 70 $\mu$m tungsten dust is ideally suited for small debris elimination. The lifetime of 60 $\mu$m diameter tungsten dust grains released at an altitude of 1100 km with inclination of 80 - 90 degrees is about 15 years.

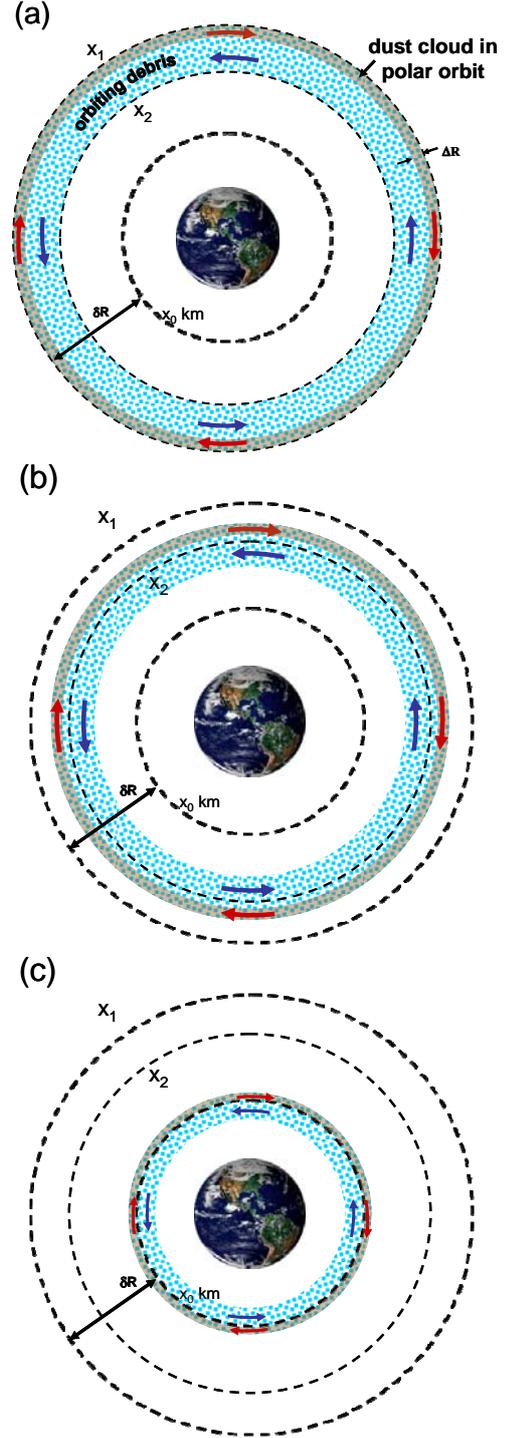

Fig. 2. Schematic illustration of the orbital dust deployment. (a) Targeted debris population is located in a shell between altitudes $X_1$ and $X_2$ (shaded blue). A torroidal dust cloud of thickness $\Delta R$ is deployed in polar orbit at the upper edge of the debris band (shaded beige) (b) Half of the debris population always counter rotates with dust. This leads to enhanced drag on the debris resulting in loss of debris altitude. The dust orbit also descends under earth's natural drag. Dust sweeps the debris population with it and leads to a "dust snow plow" effect. (c) The descents of the dust cloud and the debris population can be synchronized. Both dust and debris descends to an altitude of $x_0$ km below which earth's natural drag is sufficient to force reentry.

### 6. Concept of Operation

Based on the physics discussed above we envision releasing dust in quasi-circular orbits between 900 and 1100 km. The mass of dust required for remediation is a function of the ballistic coefficient of the orbital debris, the desired altitude reduction of the debris, and the desired altitude reduction rate. The dimension and material density of the dust grains will be optimized so that it can sustain the 'snow plow' effect. The period of induced drag on targeted small debris is purposely designed to be long (years) so that the requirement for total dust mass carried to orbit is lower. The dust cloud will spread and form a thin shell slowly spreading in azimuth with large meridianal velocity. At any given point in this shell, half of the targeted debris population will be in orbit oppositely directed to the orbiting dust mass. The interaction of dust with debris in this shell will lower the debris altitude. The dust layer itself will descend in altitude over time and in the process lower the altitude of all targeted debris from 1100 to 900 km below which the orbital lifetime of the small debris is naturally 25 years or less. Along with the debris, the injected dust will ultimately burn up in the earth's atmosphere at lower altitudes. The technique described essentially just requires the transportation of "dumb mass" (micron-sized tungsten dust) to polar orbit. No new technology development is necessary. The dust may be delivered as a secondary payload utilizing the excess capacity available in many launches going to sun synchronous orbit or as a separate dedicated dust dispensing satellite.

### 7. System Risks

Spacecraft are already designed to operate in the existing cosmic dust and orbital debris environment. The orbital debris remediation technique using tungsten dust described herein would involve higher flux than the current background. Some fraction of the dust may escape the layer and disperse in space. Certain aspects of spacecraft design are already dust impact tolerant by design. Dust grains of the size proposed by NRL will certainly not penetrate thermal blankets, spacecraft structure, or sensor baffles. Normally, earth observation satellites would point the sensors earthward and scientific satellites away from earth both nearly orthogonal to the satellite motion. Hence, the risk to satellite sensors associated with our small debris removal technique is minimal as the tungsten dust would approach in the local horizontal plane only. Solar arrays could also be degraded by dust impacts, but that effect can be mitigated by thicker cover glass. Recent laboratory tests indicate that solar cells remain unaffected by hypervelocity impact of a 100 $\mu$m glass sphere[11]. The NRL concept involves deploying a tungsten dust layer of a limited thickness, perhaps 30 to 50 km. Spacecraft could be maneuvered above or below this band using onboard propulsion and avoid the artificial dust flux altogether. Finally, tungsten dust is no longer an issue to operational spacecraft below an altitude of about 600 km because once at that altitude, the tungsten dust orbital lifetime would be brief and any interaction time with operational spacecraft below 600 km would be minimal.

### 8. Environmental Impact

Micro meteorites introduce hundreds of tons of dust in the earth's immediate environment daily[12]. Hence, injection of tens of tons of dust as required in the NRL small debris removal technique is a small perturbation to the natural dust flux into the earth's environment. Satellites are designed to perform in the natural dusty environment.

An estimate of ejecta flux resulting from hypervelocity dust/debris impacts indicates that it will be indistinguishable from the natural background of micro meteorite flux. Ignoring the small fraction of dust grain kinetic energy expended to form crater and melt the debris material, the center of mass velocity of the recoil jet, $V_R = (\sqrt{m_d/M_R})V_{rel}$, is obtained by equating the kinetic energy of the dust grain ($m_d V_{rel}^2/2$) with that of the recoil jet ($M_R V_R^2/2$) where $M_R$ is the mass of the recoil jet. The momentum of the recoil jet $P_R = M_R V_R = p_d\sqrt{M_R/m_d}$, where $p_d = m_d V_{rel}$ is the dust grain momentum in the debris frame and $M_R >> m_d$. Thus, net change in the debris momentum in a hypervelocity impact is $\Delta P_D \approx p_d\sqrt{M_R/m_d}$ so that the momentum boost factor $\kappa = \Delta P_D/p_d \approx \sqrt{M_R/m_d}$. In the inertial frame the center of mass velocity of the ejecta $V_E = V + V_R$ implies that the velocity spread with which the ejecta is emitted is $V_E - V = V_R \approx V_{rel}/\kappa$. This implies that the ejecta matter will be distributed over an altitude range $\Delta R = 2R(V_{rel}/\kappa)/V \approx 4R/\kappa$, where $V_{rel} = 2V$. For $\kappa \sim 18$ the spread of the ejecta covers a volume of $\sim 10^{21}$ m$^3$.

There are about $10^4$ pieces of small debris between 1 mm to 10 cm scale size. Let us consider the worst case scenario where all debris is of 10 cm size. Assuming a typical aluminum debris with $B = 5$, the change of debris velocity per collision at hypervelocity with a 60 $\mu$m tungsten dust grain, obtained from the conservation of momentum, is $\Delta V/V = 2\kappa(m_d/M) \sim 10^{-7}\kappa$. This implies that the reduction of debris altitude per collision $\Delta R/R = 4\kappa(m_d/M) \sim 2\times 10^{-7}\kappa$. Hence, to decrease the altitude of a debris from 1100 km to 900 km, $\sim 10^4$ collisions are required, which implies that the total number of collisions to decrease $10^4$ pieces of debris is $10^8$. Assuming that the ejecta matter ($\sim 300 m_d$) is emitted as spheres of 0.3 mg and 0.05 cm in diameter with a velocity $V_E \sim 8$ km/s, the ejecta flux will be $\sim 0.03/m^2/yr$. This is similar to comparable size micrometeorite flux [14]. A recent hypervelocity test data [15] indicates that ejecta material has a size distribution much smaller than the hypervelocity projectile which may result in ejecta flux even lower than that of the micrometeorites of comparable size.

## 9. Conclusion

We described the broad contours of a technique for eliminating small debris from the low earth orbits. A detailed description will be presented elsewhere. We studied the case where satellite fragmentations have occurred long before dust is deployed and the debris fragments have already spread around the earth. The technique is more efficient if dust can be deployed within a few revolutions of a satellite fragmentation while the debris fragments are localized in a small volume. Improvements in space situational awareness may allow this in the future. The system efficiency also depends on the value of the momentum boost factor $\kappa$. Further research is necessary to accurately determine the value of $\kappa$. In addition, more detailed orbit analysis for both dust and debris is required. The debris fragment created by collision of objects is likely to rotate. Hence the effective area exposed to the dust flow is averaged. This will result in correction to the ballistic coefficient $B = M/A$. Here we ignored this correction for simplicity.

**Acknowledgments**

This work is supported by the Office of Naval Research. Stimulating discussions with Darren McKnight and Mihaly Horanyi are acknowledged.

Certain aspects of this paper are the subject of United States Patent Application # 13/017,813.